\documentclass[a4paper]{article}
\usepackage[latin1]{inputenc}
\usepackage[T1]{fontenc}
\usepackage[english]{babel}
\usepackage{amsfonts}
\usepackage{amssymb}
\usepackage{amsmath}
\usepackage{amsthm}
\usepackage{graphicx}

\def\be{\begin{equation}}
\def\ee{\end{equation}}
\def\bc{\begin{center}}
\def\ec{\end{center}}

\begin{document}

\title{On the equivalence of Hopfield Networks and Boltzmann Machines}


\author{Adriano Barra\footnote{Dipartimento di Fisica, Sapienza Universit\`a di Roma.},
Alberto Bernacchia \footnote{Department of Neurobiology, Yale University.}, \ Enrica Santucci \footnote{Dipartimento di Matematica, Universit\`a degli Studi dell'Aquila.} \ and Pierluigi Contucci \footnote{Dipartimento di  Matematica,
    Alma Mater Studiorum Universit\`a di Bologna.}}

\date{January 2012}

\maketitle

\begin{abstract}
A specific type of neural network, the Restricted Boltzmann Machine (RBM), is implemented for classification and feature detection in machine learning.
RBM is characterized by separate layers of visible and hidden units, which are able to learn efficiently a generative model of the observed data.
We study a "hybrid" version of RBM's, in which hidden units are analog and visible units are binary, and we show that thermodynamics of visible units are equivalent to those of a Hopfield network, in which the $N$ visible units are the neurons and the $P$ hidden units are the learned patterns.
\newline
We apply the method of stochastic stability to derive the thermodynamics of the model, by considering a formal extension of this technique to the case of multiple sets of stored patterns, which may act as a benchmark for the study of correlated sets.
\newline
Our results imply that simulating the dynamics of a Hopfield network, requiring the update of $N$ neurons and the storage of $N(N-1)/2$ synapses, can be accomplished by a hybrid Boltzmann Machine, requiring the update of $N+P$ neurons but the storage of only $NP$ synapses.
In addition, the well known glass transition of the Hopfield network has a counterpart in the Boltzmann Machine: It corresponds to an optimum criterion for selecting the relative sizes of the hidden and visible layers, resolving the trade-off between flexibility and generality of the model.
The low storage phase of the Hopfield model corresponds to few hidden units and hence a overly constrained RBM, while the spin-glass phase (too many hidden units) corresponds to unconstrained RBM prone to overfitting of the observed data.

\end{abstract}

\section{Introduction}

A common goal in Machine Learning is to design a device able to reproduce a given system, namely to estimate the probability distribution of its possible states \cite{AI}. 
When a satisfactory model of the system is not available, and its underlying principles are not known, this goal can be achieved by the observation of a large number of samples \cite{peter}.
A well studied example is the visual world, the problem of estimating the probability of all possible visual stimuli \cite{pitkov}.
A fundamental ability for the survival of living organisms is to predict which stimuli will be encountered and which are more or less likely to occur. 
On this purpose, the brain is believed to develop an internal model of the visual world, to estimate the probability and respond to the occurrence of various events \cite{hotel1},\cite{hotel2}.

Ising-type neural networks have been widely used as generative models of simple systems \cite{hertz},\cite{BarraCW}.
Those models update the synaptic weights between neurons according to a specific learning rule, depending on the neural activity driven by a given set of observations; after learning, the network is able to generate a sequence of states whose probabilities match those of the observations.
Popular examples of Ising models, characterized by a quadratic energy function and a Boltzmann distribution of states, are the Hopfield model \cite{amit}\cite{hopfield} and Boltzmann Machines (BM) \cite{hinton1}.
Boltzmann Machines (BM) have been designed to capture the complex statistics of arbitrary systems by dividing neurons in two subsets, visible and hidden units: marginalizing the Boltzmann distribution over the hidden units allows the BM to reproduce, through the visible units, arbitrarily complex distributions of states, by learning the appropriate synaptic weights \cite{hinton1}.
State of the art feature detectors and classifiers implement a specific type of BM, the Restricted Boltzmann Machine (RBM), because of its efficient learning algorithms \cite{benjo}.
The RBM is characterized by a bipartite topology in which hidden and visible units are coupled, but there is no interaction within either set of visible or hidden units \cite{hinton2}.

\begin{figure}[tb] \begin{center}
\includegraphics[width=.28\textwidth]{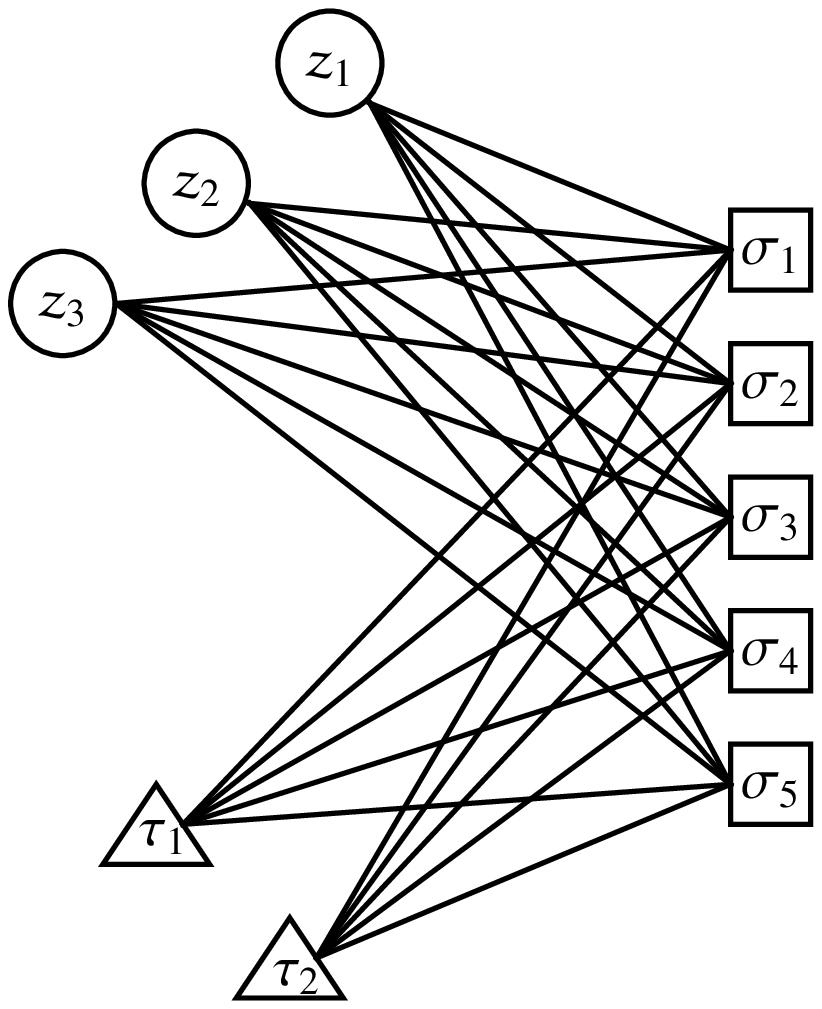} \ \ \ \ \ \ \ \ \ \ \ \ \ \ \
\includegraphics[width=.28\textwidth]{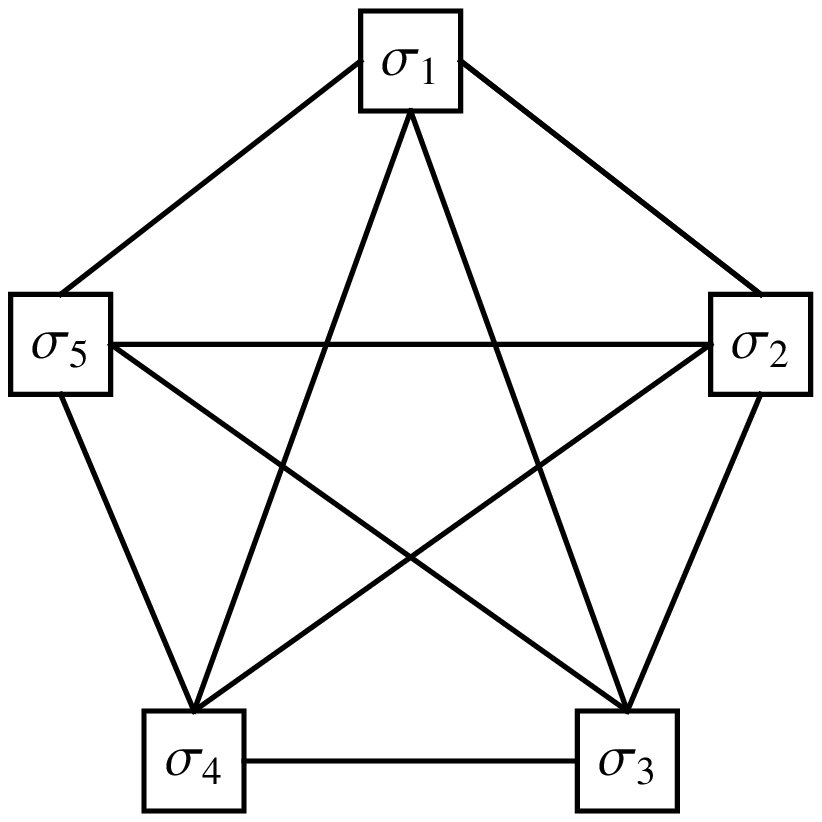}
\caption{\label{fig:distr} Left panel: Schematic representation of a Hybrid Boltzmann Machine (HBM) where the hidden units are analog ($z,\tau$ variables) and the visible units are digital ($\sigma$ variables).
The two sets of hidden units, $z$ and $\tau$, represent two feature sets that are both connected to the layer of visible units $\sigma$.
The layers of hidden and visible units are reciprocally connected, but there are no intra-layer connections, thus forming a bipartite topology.
Right panel: Schematic representation of the equivalent Hopfield neural network built upon the visible units only, with an internal fully connected structure.}
\end{center}
\end{figure}

All neurons of RBM's are binary, both the visible and the hidden units.
The analog equivalent of RBM, the Restricted Diffusion Networks, have all analog units and have been described in \cite{ristretto}\cite{benjo}.
Here we study the case of a "hybrid" Boltzmann Machine (HBM), in which the hidden units are analog and the visible units are binary (Fig.$1$ left).
We show that the HBM, when marginalized over the hidden units, is equivalent to a Hopfield network (Fig.$1$ right), where the $N$ visible units are the neurons and the $P$ hidden units are the learned patterns.
Although the Hopfield network can generate probability distributions in a limited space, it has been widely studied for its associative and retrieval properties.
The exact mapping proven here introduces a new way to simulate Hopfield networks, and allows a novel interpretation of the spin glass transition, which translates into an optimal criterion for selecting the relative size of the hidden and visible layers in the HBM.

We use the method of stochastic stability to study the thermodynamics of the system in the case of analog synapses.
This method has been previously described in \cite{ac}, \cite{BGG1}, and offers an alternative approach to the replica trick for studying Ising-type neural networks, including the Hopfield model and the HBM.
We analyze the model with two non-interacting sets of hidden units in the HBM, which corresponds to two sets of uncorrelated patterns in the Hopfield network, and study the thermodynamics with the assumption of replica symmetry.
We extend the theory to cope with two sets of interconnected hidden layers, coresponding to sets of correlated patterns, and we show that their interaction acts as a noise source for retrieval.

\section{Statistical equivalence of HBM and Hopfield networks}
We define a "hybrid" Boltzmann Machine (HBM, see Fig. $1$ left) as a network in which the activity of units in the visible layer is discrete, $\sigma_i=\pm1, \ i \in (1,...,N)$ (digital layer), and the activity in the hidden layer is continuous (analog layer).
The layers of hidden and visible units are reciprocally connected, but there are no intra-layer connections, thus forming a bipartite topology.
We assume that the layer of hidden units is further divided into two sets, both described by continuous variables, $z_\mu, \tau_{\nu} \in\Re, \ \mu \in (1,..,P), \ \nu \in (1,...,K)$.
We will consider the case of interacting hidden units (connections between $z$ and $\tau$) in the next Section.
In order to maintain a parsimonious notation, in this section we consider a single hidden layer, e.g. only the layer defined by the variables $z$.

The synaptic connections between units in the two layers are fixed and symmetric, and are defined by the synaptic matrix $\xi_i^\mu$.
The input to unit $\sigma_i$ in the visible (digital) layer is the sum of the activities in the hidden (analog) layer weighted by the synaptic matrix, i.e. $\sum_\mu\xi_i^\mu z_\mu$.
The input to unit $z_\mu$ in the hidden (analog) layer is the sum of the activities in the visible (digital) layer, weighted by the synaptic matrix, i.e. $\sum_i\xi_i^\mu\sigma_i$.
In the following, we denote by $z$ the set of all hidden $\{z_\mu\}$ variables, and by $\sigma$ the set of all visible $\{\sigma_i\}$ variables.

The dynamics of the activity is different in the two layers; in the analog layer it changes continuously in time, while in the digital layer it changes in discrete steps.
The activity in the hidden (analog) layer follows the stochastic differential equation
\begin{equation}
\label{langevin}
T\frac{dz_\mu}{dt} = - z_\mu(t)+\sum_i\xi_i^\mu\sigma_i+\sqrt{\frac{2T}{\beta}}\;\zeta_\mu(t),
\end{equation}
where $\zeta$ is a white gaussian noise with zero mean and covariance $\left<\zeta_\mu(t)\zeta_\nu(t')\right>=\delta_{\mu\nu}\;\delta(t-t')$.
The parameter $T$ quantifies the timescale of the dynamics, and the parameter $\beta$ determines the strength of the fluctuations.
The first term in the right hand side is a leakage term, the second term is the input signal and the third term is a noise source.
Since noise is uncorrelated between different hidden units, they evolve independently.
Eq.(\ref{langevin}) describes an Ornstein-Uhlembeck diffusion process \cite{tuckwell} and, for fixed values of $\sigma$, the equilibrium distribution of $z_\mu$ is a Gaussian distribution centered around the input signal, which is equal to

\begin{equation}
\label{cond1}
Pr(z_\mu|\sigma)=\sqrt{\frac{\beta}{2\pi}}\exp\left[-\frac{\beta}{2} \Big(z_\mu-\sum_{i}\xi_i^\mu\sigma_i\Big)^2\right]
\end{equation}
In order for this equilibrium distribution to hold, the activity of digital units $\sigma$ must be constant, while in fact it depends on time.
However, we assume that the timescale of diffusion $T$ is much faster than the rate at which the digital units are updated.
Therefore, a different equilibrium distribution for $z$, characterized by different values of $\sigma$, holds between each subsequent update of $\sigma$.
Since hidden units are independent, their joint distribution is the product of individual distributions, i.e. $Pr(z|\sigma)=\prod_{\mu=1}^PPr(z_\mu|\sigma)$.

The activity in the visible (digital) layer follows a standard Glauber dynamics for Ising-type systems \cite{amit}.
At a specified sequence of time intervals (much larger than $T$), the activity of units in the digital layer is updated randomly according to a probability that depends on their input. 
While updating the digital units $\sigma$, the analog variables $z$ are fixed, namely the update of digital units is instantaneous.
The activity of a unit $\sigma_i$ is independent on other units, and the probability is a logistic function of its input, i.e.

\begin{equation}
\label{glauber}
Pr(\sigma_i|z)=\frac{\exp[\beta \sigma_i\sum_\mu\xi_i^\mu z_\mu]}{\exp[\beta \sum_\mu\xi_i^\mu z_\mu]+\exp[-\beta \sum_\mu\xi_i^\mu z_\mu]}
\end{equation}
Note that this distribution is normalized, namely $Pr(\sigma_i=+1|z)+Pr(\sigma_i=-1|z)=1$.
Since visible units are independent, their joint distribution is the product of individual distributions, i.e. $Pr(\sigma|z)=\prod_{i=1}^NPr(\sigma_i|z)$.

Given the conditional distributions of either layers, Eqs.(\ref{cond1},\ref{glauber}), we can determine their joint distribution, $Pr(\sigma,z)$, and the marginal distributions $Pr(z)$ and $Pr(\sigma)$, apart from a normalization factor.
We use Bayes' rule, $Pr(\sigma,z)=Pr(z|\sigma)Pr(\sigma)=Pr(\sigma|z)Pr(z)$, and we use the fact that marginal distributions depend on single layer variables.
The result is, for the joint distribution 

\begin{equation}
\label{seiuncane}
Pr(\sigma,z)\propto\exp\left(-\frac{\beta}{2}\sum_\mu z_\mu^2 +\beta\sum_{i,\mu}\sigma_i\xi_i^\mu z_\mu \right).
\end{equation}
The marginal distribution of the visible units is equal to

\begin{equation}
Pr(\sigma)\propto\exp\left[\frac{\beta}{2}\sum_{i,j}\left(\sum_\mu\xi_i^\mu\xi_j^\mu\right)\sigma_i\sigma_j\right]
\end{equation}
As explained in more detail in the next section, this probability distribution is equal to the distribution of a Hopfield network, where the synaptic weights of the Hopfield network are given by the expression in round brackets.
The stored patterns of the Hopfield model corresponds to the synaptic weights of the HBM, described by the $\xi$ variables, and the number of patterns corresponds to the number $P$ of hidden units.

Therefore, we have shown that the HBM and Hopfield network admit the same probability distribution, once the hidden variables of the HBM are marginalized, and the HBM and Hopfield network are statistically equivalent.
In other words, a configuration $\sigma$ in the Hopfield network has the same probability as the same configuration $\sigma$ in the HBM, when averaged over the hidden configurations $z$.
Retrieval in the Hopfield network corresponds to the case in which the HBM learns to reproduce a specific pattern of neural activation.
The maximum number of patterns $P$ that can be retrieved in a Hopfield network is known \cite{amit}, and is equal to $0.14\cdot N$.
If the number of patterns exceeds this limit, the network is not able to retrieve any of them.
On the other hand, if the HBM has a very large number $P$ of hidden variables, this provokes over-fitting in learning the observed patterns, and the HBM is not able to reproduce the statistics of the observed system.
The correspondence between Hopfield network and HBM allows recognizing that the maximum number of hidden variables in the HBM is $0.14\cdot N$.

We check this prediction by numerical simulations of the HBM.
We pick each element of the synaptic matrix $\xi_i^\mu$ independently from a Bernouilli distribution, $\xi_i^\mu=1/\sqrt{N}$ or $\xi_i^\mu=-1/\sqrt{N}$ with $50\%$ probability (the scaling with $N$ is imposed for comparison to the original Hopfield model).
We set the number of neurons in the visible layer as $N=1000$, the timescale of dynamics of hidden units is $T=1$, and we use $T$ as a reference time unit. In each simulation, we update the visible units every $1T$ and we run the simulation for $1000T$, therefore performing one thousand updates of the visible units.
We simulate the dynamics of hidden units by standard numerical methods for stochastic differential equations and using a time step of 0.01T.
In different simulations we vary the values of the noise amplitude by manipulating $\beta =0.5,2,10$, and the number of hidden units $P=50,100,150,200$.
We observe the overlap of the activity of visible units with each one of the pattern $\mu$ by computing $\sum_i\xi_i^\mu\sigma_i/\sqrt{N}$, such that overlap equal to one for some value of $\mu$ implies that visible units precisely align to that pattern $\mu$.
In each simulation, we initialize the hidden units at random and the visible units exactly aligned to one of the patterns.

Fig.2 shows the results of simulations, the dynamics of the overlap of visible units with all patterns for different values of the parameters $\beta$ and $P$.
For high noise, $\beta=0.5$, no retrieval is possible and all overlaps are near zero regardless of the number of hidden units $P$.
For intermediate noise, $\beta=2$, retrieval is possible provided that the number of hidden units is not too large.
The prediction of the Hopfield network is that retrieval is lost at about $P\simeq 0.06N=60$ \cite{amit}, accurately matching our findings.
For low noise, $\beta=10$, retrieval is maintained up to large values of $P$.
In the low noise regime, the Hopfield network can retrieve a number of patterns near its maximum, i.e. $P=0.14N=140$, which again matches well with the results of our simulations.

\begin{figure}[tb] \begin{center}
\includegraphics[width=0.7\textwidth]{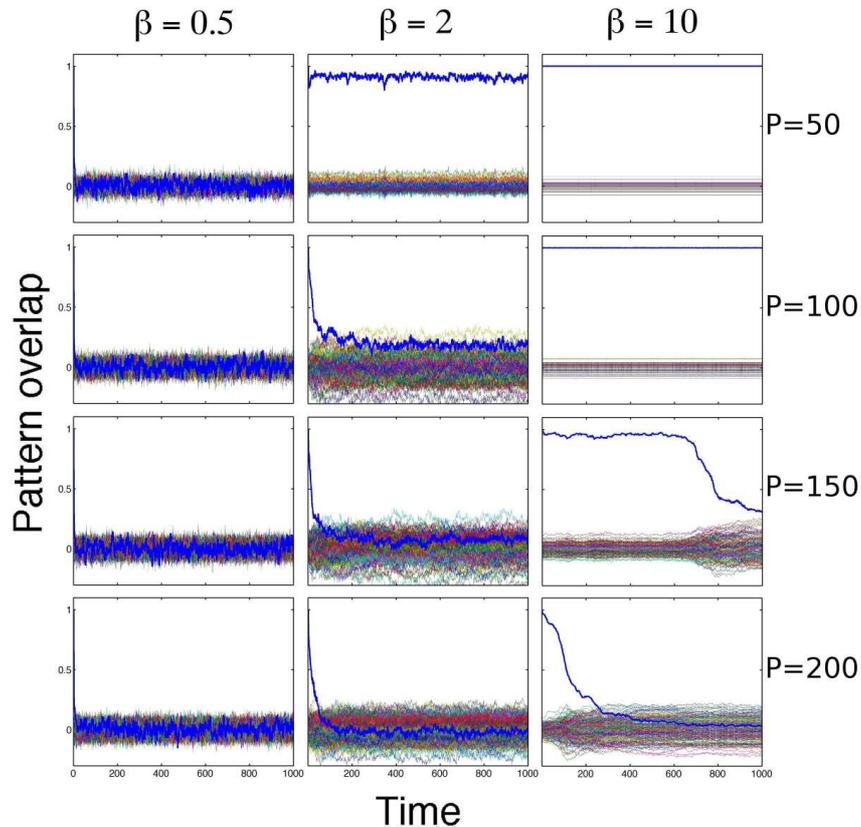}
\caption{\label{fig:simres}
Dynamics of the overlap of visible units with all patterns for different values of the parameters $\beta$ and $P$.
Simulations run for $1000$ units of time, which corresponds to $1000$ updates of the visible units.
Thick blue line in each panel shows the overlap of visible units with the pattern imposed by the initial condition, other lines show the overlaps with all other patterns.
No retrieval is observed (overlap$\sim0$) for high noise, $\beta=0.5$, while for intermediate $\beta=2$ and low noise $\beta=10$ retrieval is possible (overlap$\sim1$) for a small number of hidden units (patterns) $P$.
Results of simulations match with the theory of Hopfield networks.
} 
\end{center}
\end{figure}

\section{Thermodynamic theory of HBM}
In canonical statistical mechanics, a system is described by the probability distribution of each one of its possible states. 
In the HBM, a given state is associated with its probability according to the Boltzmann distribution.
This distribution is expressed by Eq.(\ref{seiuncane}), which we rewrite while reintroducing the variables $\tau$ dropped in previous section and by defining the Hamiltonian function 
\begin{equation}\label{hamilton1}
H_{hbm}(\sigma,z,\tau;\xi,\eta) = \frac{1}{2}\left(\sum_\mu z_\mu^2+\sum_\nu\tau_\nu^2\right)-\sum_{i}\sigma_i\left(\sum_\mu\xi_i^\mu z_\mu+\sum_\nu\eta_i^\nu\tau_\nu\right),
\end{equation}
We denote by $\xi$ the set of all $\{\xi_i^\mu\}$ variables, and by $\eta$ the set of all $\{\eta_i^\nu\}$ variables, where $\eta_i^\nu$ is the synaptic matrix for the connections with the $\tau$ layer.
The Boltzmann distribution depends on the parameters $\beta,\xi,\eta$, and its expression includes the normalization factor $Z$: 

\begin{equation}
\label{boltzdist}
Pr(\sigma,z,\tau)=\exp\left[-\beta H_{hbm}(\sigma,z,\tau;\xi,\eta)\right]Z(\beta,\xi,\eta)^{-1}
\end{equation}
The partition function $Z$ corresponds to the normalization factor of the Boltzmann distribution, and is defined as
\begin{equation}
\label{partitionz}
Z(\beta,\xi,\eta)=\sum_{\sigma}\int\prod_{\mu=1}^{P}dz_{\mu}\int\prod_{\nu=1}^{K}d\tau_{\nu}\;\exp{(-\beta H_{hbm}(\sigma,z,\tau;\xi,\eta))},
\end{equation}
Using this definition of the partition function, it is straightforward to show that the Bolzmann distribution, Eq.(\ref{boltzdist}), is normalized.
In order to marginalize the hidden variables, we use the following identity, the Gaussian integral:

\begin{equation}\label{gauss}
\int_{-\infty}^{+\infty}dz\exp{\left[-\beta \left(\frac{z^2}{2}-az\right)\right]} = \sqrt{\frac{2\pi}{\beta}}\exp{\left(\beta\frac{a^2}{2}\right)}.
\end{equation}
Using this identity, we marginalize the analog variables $z$ and $\tau$ in Eq.(\ref{partitionz}), and we obtain

\begin{equation}
\label{parthopf}
Z(\beta,\xi,\eta)=\left(\frac{2\pi}{\beta}\right)^{\frac{P+K}{2}}\sum_{\sigma}\exp(- \beta H_{hop}(\sigma;\xi,\eta)),
\end{equation}
where we define the following Hamiltonian:
\begin{equation}\label{monopartito}
H_{hop}(\sigma;\xi,\eta)=-\frac{1}{2}\sum_{i,j}^{N} \left(\sum_{\mu=1}^{P}\xi_{i}^{\mu}\xi_{j}^{\mu}+\sum_{\nu=1}^{K}\eta_{i}^{\nu}\eta_{j}^{\nu}\right) \sigma_{i}\sigma_{j}.
\end{equation}
This is the Hamiltonian of a Hopfield neural network.
This result connects the two Hamiltonians of the Hopfield network and the Boltzmann Machine and states that thermodynamics obtained by the first cost function, Eq.(\ref{hamilton1}), is the same as the one obtained by the second one, Eq.(\ref{monopartito}).
This offers a connection between retrieval through free energy minimization in the Hopfield network and learning through log-likelihood estimation in the HBM \cite{amit}\cite{benjo}.
Note that observable quantities stemmed from HBM are equivalent in distribution, and not pointwise, to the corresponding ones in the Hopfield network.

Next, we calculate the free energy, which allows determining the value of all relevant quantities and the different phases of the system.
The thermodynamic approach consists in averaging all observable quantities over both the noise and the configurations of the system.
Therefore, we define two different types of averaging, the average $\omega$ over the state configurations $\sigma, z, \tau$, and the average $\mathbb{E}$ over the synaptic weights (quenched noise) $\xi, \eta$.
Note that a given HBM is defined by a fixed and constant value of the synaptic weights $\xi, \eta$.
However, those synaptic weights are taken at random from a given distribution, and different realizations of the synaptic weights correspond to different HBM's.
Since we are interested in determining the average behavior of a "typical" HBM, we average the relevant quantities over the distribution of synaptic weights.

The average $\omega$ of a given observable $O$ under the Boltzmann distribution is defined as
\begin{equation}
\omega(O)=Z(\beta;\xi,\eta)^{-1}\sum_{\sigma}\int\prod_{\mu=1}^{P}dz_{\mu}\int\prod_{\nu=1}^{K}d\tau_{\nu}\;O(\sigma,z,\tau)\exp{(-\beta H_{hbm}(\sigma,z,\tau;\xi,\eta))}.
\end{equation}
The average $\mathbb{E}$ of a given observable $F$ over the distribution of synaptic weights is defined as

\begin{equation}
\mathbb{E}[F(\xi,\eta)]=\int d\mu(\xi)\int d\mu(\eta)F(\xi,\eta),
\end{equation}
where $\mu$ is the standard Gaussian measure, $d\mu(\xi)=d\xi \exp(-\xi^2/2)/\sqrt{2\pi}$.
Note that the standard Hopfield network is built with random binary patterns $\xi$, while we use Gaussian patterns here: Despite retrieval with the former choice has been extensively studied, we have chosen the latter in order to show a novel technique, stochastic stability, for studying the related thermodynamics.  
For finite $N$, this is known to be equivalent to the former case, despite for infinite neurons a complete picture of the quality of the retrieval is still under discussion.

We define the free energy as
\begin{equation}
A(\beta)=\frac{1}{N}\mathbb{E}\left[\log Z(\beta;\xi,\eta)\right].
\end{equation}
Since the free energy is proportional to the logarithm of the partition function, and due to the additive properties of the logarithm, $\log(A\cdot B)=\log A + \log B$, we neglect the factor $(2\pi/\beta)^{(P+K)/2}$ in $Z(\beta; \xi, \eta)$ (see Eq. (\ref{parthopf})), as it gives a negligible contribution to the free energy in the thermodynamic limit. 
We also neglect the factor in the Bolzmann average $\omega$ as it appears both at the numerator and denominator and therefore it cancels out. 
Thermodynamics can be described by the standard Gaussian measure.

In the HBM, parameters $P$ and $K$ determine the number of neurons in the hidden layers, while in the Hopfield model they represent the number of patterns stored in the network, or the number of stable states that can be retrieved.
We consider the "high storage" regime, in which the number of stored patterns is linearly increasing with the number of neurons \cite{amit}.
In HBM, this corresponds to the case in which the sizes of the hidden and visible layers are comparable.
Their relative size is quantified by defining two control parameters $\alpha,\gamma \in \mathbb{R}^+$ as
\begin{equation}
\alpha=\lim_{N\rightarrow\infty}\frac{P}{N} \ \ \ \ \ \ \ \gamma=\lim_{N\rightarrow\infty}\frac{K}{N}.
\end{equation}
We further introduce the order parameters $q,p,r$, called overlaps, as
\begin{equation}
q_{ab}=\frac{1}{N}\sum_{i=1}^{N}\sigma^{a}_{i}\sigma^{b}_{i},\ \ \ \ \ \ \ p_{ab}=\frac{1}{P}\sum_{\mu=1}^{P}z^{a}_{\mu}z^{b}_{\mu},\ \ \ \ \ \ \
r_{ab}=\frac{1}{K}\sum_{\nu=1}^{K}\tau^{a}_{\nu}\tau^{b}_{\nu}.
\end{equation}
These objects describe the correlations between two different realizations  of the system (two different replicas $a,b$).
We also define the averages of these overlaps with respect to both state configurations and synaptic weights (quenched noise).
Since the overlaps involve two realizations of the system ($\sigma^{a}, \sigma^{b}$), the Boltzmann average is performed over both configurations.
With some abuse of notation, we use the symbol $\omega$ to also represent the Boltzmann average over two-system configurations.
Therefore, the average overlaps are defined as

\begin{equation}
\bar{q}=\mathbb{E}\;\omega(q_{ab}), \ \ \ \ \ \ \ \bar{p}=\mathbb{E}\;\omega(p_{ab}), \ \ \ \ \ \ \ \bar{r}=\mathbb{E}\;\omega(r_{ab}).
\end{equation}
The goal of next section is to find an expression for the free energy in terms of these order parameters.
While all configurations of the system are possible, only a subset of them has a significant probability to occur.
In canonical thermodynamics, those states are described by the minima of the free energy with respect to the order parameters.
The free energy is the difference among the energy and the entropy, and its minimization corresponds to energy minimization and entropy maximization.

\subsection{Multiple-layer stochastic stability}

By definition of HBM, we assume that no external field acts on the network; inputs to all neurons are generated internally by other neurons.
The overall stimulus felt by an element of a given layer is the sum, synaptically weighted, of the activity of neurons in  the other layers.
Note that neurons are connected in loops, because a neuron receiving input from a layer also projects back to same layer.
Therefore, the HBM is a recurrent network, and this makes the calculation of the free energy complicated.
However, the free energy can be calculated in specific cases by using a novel technique that has been developed in \cite{BGG1}, which extends the stochastic stability developed for the analysis of spin glasses \cite{ac}.
This technique introduces an external field acting on the system which "imitates" the internal, recurrently generated input, by reproducing its average statistics.
While the external, fictitious input does not reproduce the statistics of order two and higher, it represent correctly the averages.
These external inputs are denoted as $\tilde{\eta}$ (one for each neuron in each layer) and are distributed following the standard Gaussian distribution $\mathcal{N}[0,1]$.

In order to recover the second order statistics, the free energy is interpolated smoothly between the case in which all inputs are external, and all high order statistics is missing, and the case in which all inputs are internal, describing the original HBM.
We use the interpolating parameter $t \in [0,1]$, such that for $t=0$ the inputs are all external and the calculation straightforward, while for $t=1$ the full HBM is recovered.

Therefore, we define the interpolating free energy as
\begin{eqnarray}\label{interpolazione}
\tilde{A}(\beta,t)&=&\frac{1}{N}\mathbb{E}\log\sum_{\sigma}\int\prod_{\mu=1}^{P}dz_{\mu}
\int\prod_{\nu=1}^{K}d\tau_{\nu}\exp{\Big[-\frac{\beta}{2}\Big(\sum_{\mu=1}^{P}z_{\mu}^{2}+\sum_{\nu=1}^{K}\tau_{\nu}^{2}\Big)\Big]}\\
&\cdot&\exp{\sqrt{t}\Big(\beta \sum_{i,\mu}\sigma_i\xi_i^\mu z_\mu -\sum_{i,\nu}\sigma_i\eta_i^\nu \tau_\nu\Big)}\nonumber\\
&\cdot&\exp{\sqrt{1-t}\Big(a\sum_{i=1}^{N}\tilde{\eta}_{i}\sigma_{i}+b\sum_{\mu=1}^{P}\tilde{\eta}_{\mu}z_{\mu}+
c\sum_{\nu=1}^{K}\tilde{\eta}_{\nu}\tau_{\nu}\Big)}\exp{\Big[(1-t)\Big(\frac{h}{2}\sum_{\mu=1}^{P}z_{\mu}^{2}+
\frac{\epsilon}{2}\sum_{\nu=1}^{K}\tau_{\nu}^{2}\Big)\Big]}.\nonumber
\end{eqnarray}
In addition to the fictitious fields $\tilde{\eta}$, we have introduced the auxiliary parameters $a, b, c$, which serve to weight the external fields.
We also introduced an additional leakage (second order) term, parameterized by $h$ and $\epsilon$.
Those parameters are chosen once for all in Appendix $1$ and $2$ in order to separate the contribution of mean and fluctuations of the order parameters in the final expression of the free energy.
This technique is called multiple layer stochastic stability because each of the three layers are perturbed by external fictitious inputs to simplify the expression of the free energy.

The free energy at $t=0$ is characterized by one-body terms and is calculated in Appendix $1$.
The result is Eq.(\ref{interpo0}) and is equal to

\begin{eqnarray}
\label{cristo}
\tilde{A}(\beta,t=0)&=&\log2+\int d\mu(\eta)\log\cosh(\sqrt{\beta(\alpha\bar{p}+\gamma\bar{r})}\eta)+\frac{\alpha+\gamma}{2}\log(1-\beta(1-\bar{q}))^{-1}\nonumber\\
&+&\frac{\beta(\alpha+\gamma)}{2}\frac{\bar{q}}{1-\beta(1-\bar{q})}.
\end{eqnarray}
In order to derive the expression of the free energy for the HBM, namely for $t=1$, we use the sum rule

\begin{equation}
\label{porcodio}
\tilde{A}(\beta,t=1)=\tilde{A}(\beta,t=0)+\int_{0}^{1}dt^{'}\Big(\frac{d\tilde{A}(\beta,t)}{dt}\Big)_{t=t^{'}}.
\end{equation}
Therefore, we need to compute the derivative of the interpolating free energy in order to recover the free energy of the HBM ($t=1$).
We calculate the derivative in Appendix $2$ (Eq.(\ref{strilla})), and the result is

\begin{equation}
\label{seiuncojone}
\frac{d\tilde{A}}{dt} = S(\alpha,\beta,\gamma) +  \frac{\beta}{2}(\bar{q}-1)(\alpha\bar{p}+\gamma\bar{r})-\frac{\beta(\alpha+\gamma)}{2},
\end{equation}
where the function $S$ is the source of the fluctuations of the order parameters, and is equal to

\begin{equation}
\label{diacane}
S(\alpha,\beta,\gamma)  = -\frac{\beta}{2}\langle(q_{12}-\bar{q})[\alpha(p_{12}-\bar{p})
+\gamma(r_{12}-\bar{r})]\rangle.
\end{equation}
In the following, we neglect the contribution of fluctuations, therefore we set $S=0$.
The integral in Eq.(\ref{porcodio}) is calculated by substituting the derivative in Eq.(\ref{seiuncojone}) with $S=0$ and, since the latter does not depend on $t$, it can be integrated simply multiplying by one.
Further, we substitute the expression of the free energy at $t=0$, Eq.(\ref{cristo}), and we obtain the final expression for the free energy of the HBM ($t=1$).
The resulting expression is called $A^{RS}$, since it does not include fluctuations of the overlaps, and this corresponds to the replica symmetric ($RS$) solution in statistical mechanics.

\begin{eqnarray}
\label{cocaina}
A^{RS}&=&\log2+\int d\mu(\eta)\log\cosh(\sqrt{\beta(\alpha\bar{p}+\gamma\bar{r})}\eta)+\frac{\alpha+\gamma}{2}\log(1-\beta(1-\bar{q}))^{-1}\\
&+& \frac{\beta(\alpha+\gamma)}{2}\frac{\bar{q}}{1-\beta(1-\bar{q})}+ \beta(\bar{q}-1)(\alpha\bar{p}+\gamma\bar{r})/2 - \beta(\alpha+\gamma)/2. \nonumber
\end{eqnarray}
In Appendix $3$, we derive the free energy in the case in which an additional external input is applied to the HBM, in order to force the retrieval of stored patterns.
In the next section, we minimize the free energy with respect to the order parameters, in order to study the phases of the system.

\subsection{Free energy minimization and phase transition}

We minimize the free energy (\ref{cocaina}) with respect to the order parameters $\bar{q},\bar{p},\bar{r}$.
This is accomplished by imposing the following equations
\begin{equation}\nonumber
\partial_{\bar{q}}A^{RS}=0,\qquad\partial_{\bar{p}}A^{RS}=0,
\qquad\partial_{\bar{r}}A^{RS}=0.
\end{equation}
This gives the following system of integro-differential equations to be simultaneously satisfied
\begin{align}
&\partial_{\bar{q}}A^{RS}=\frac{\beta}{2}\Big(\alpha\bar{p}+\gamma\bar{r}-\frac{(\alpha+\gamma)
\bar{q}\beta}{(1-\beta(1-\bar{q}))^{2}}\Big)=0,\\
&\partial_{\bar{p}}A^{RS}=\frac{\alpha\beta}{2}\Big(\bar{q}-\int d\mu(\eta)\tanh^{2}\Big(\eta\sqrt{\beta(\alpha\bar{p}+\gamma\bar{r})}\Big)=0, \\
&\partial_{\bar{r}}A^{RS}=\frac{\gamma\beta}{2}\Big(\bar{q}-\int d\mu(\eta)\tanh^{2}\Big(\eta\sqrt{\beta(\alpha\bar{p}+\gamma\bar{r})}\Big)=0,
\end{align}
Note that, since the two hidden layers act symmetrically on the visible layer, in the sense that the synaptic weights are distributed identically, one of the above equations is redundant and the minimization condition is summarized by the following two equations
\begin{eqnarray}
\alpha\bar{p}+\gamma\bar{r}&=&\frac{\bar{q}(\alpha+\gamma)\beta}{(1-\beta(1-\bar{q}))^2},\\ \label{crack}
\bar{q}&=&\int d\mu(\eta)\tanh^{2}\left(\frac{\beta \sqrt{(\alpha+\gamma)\bar{q}}\eta}{1-\beta(1-\bar{q})}\right).
\end{eqnarray}
These equations describe a minimum of the free energy, as can be checked by calculating the second-order derivatives of the free energy and verifying that the Hessian has a positive determinant.
The minima of free energy in the case of imposed retrieval are discussed in Appendix $3$.

Next, we study the phase transitions of the system by looking at divergences of the rescaled order parameters.
If the overlap $\bar{q}$ is zero, then all neurons in the visible layer are uncorrelated, implying that all neurons have random activity and the system has no structure.
The value of parameters for which the transition to $\bar{q}=0$ occurs corresponds to the case in which the fluctuations of $\sqrt{N}q$ diverge, and this defines the critical region.
To evaluate the critical region, we study for which values of the parameters $\alpha,\beta,\gamma$ the squared order parameter $N\bar{q}^2$ diverges.
This is obtained by expanding the hyperbolic tangent in Eq.(\ref{crack}) to the second order, which gives a meromorphic expression for the overlap.
This expression diverges at the critical region, which is characterized by
\begin{equation}
\beta = \frac{1}{1+ \sqrt{\alpha + \gamma}}.
\end{equation}

The above equations are consistent with and generalize the results obtained in \cite{amit}.
Since the hidden layers are not connected, and $z,\tau$ are conditionally independent, they are equivalent to a single hidden layer of $P+K$ neurons.
Therefore, the equivalent Hopfield network stores $P+K$ independent patterns.
The case of interacting (correlated) patterns is studied in the next section.

\subsection{Analysis of interacting hidden layers}

In this section we study the case in which the two hidden layers are connected by mild interactions.
When the hidden units in the two separate layers interact, the performance of the network may change.
We study this case for small interaction strengths, within a mean field approximation, in order to be able to obtain approximate results.
We show that the two hidden layers act reciprocally as an additional noise source affecting the retrieval of stored patterns in the visible layer, i.e. the retrieval of the activities $\sigma$.

We introduce the interacting energy of the HBM, denoted by $H_I$, where $I$ stands for "interacting layers":
\begin{equation}\label{inter}
H_I(\sigma,z,\tau; \xi,\eta)=\frac{-1}{\sqrt{N}}\sum_{i \mu} \xi_i^{\mu}\sigma_i z_{\mu} + \frac{-1}{\sqrt{N}}\sum_{i k}\xi_i^k\sigma_i \tau_k + \frac{-\epsilon}{\sqrt{N}} \sum_{\mu k} \xi_{\mu k} z_{\mu} \tau_k,
\end{equation}
where the last term accounts for the interaction between hidden layers, and its strength is controlled by the parameter $\epsilon$, which is assumed to be small.

The rigorous analysis of this model is complicated and still under investigation \cite{tantari}.
However, for small $\epsilon$, exact bounds can be obtained by first-order expansion.
We will proceed as follows: first we marginalize over one layer (either $\tau$ or $z$) and we find an expression of the interacting partition function depending on the two remaining ones. 
Then, because of the symmetry between the hidden layers, we perform the same operation marginalizing the interacting partition function with respect to the other hidden layer. 
Last, we sum the two expression and divide the result by two: This should represent the average behavior of the neural network, whose properties are then discussed.

The interacting partition function $Z_I$, associated to the energy (\ref{inter}), can be written as
\begin{equation}
Z_I= \sum_{\sigma}\int \prod_{{\mu}=1}^P d \mu(z_{\mu})\int \prod_{{\nu}=1}^K d \nu(\tau_{\nu})\exp\left( -\beta \left( \frac{-1}{\sqrt{N}}\sum_{i \mu} \xi_i^{\mu}\sigma_i z_{\mu} + \frac{-1}{\sqrt{N}}\sum_{i \nu}\xi_i^{\nu}\sigma_i \tau_{\nu} + \frac{-\epsilon}{\sqrt{N}} \sum_{\mu \nu} \xi_{\mu \nu} z_{\mu} \tau_{\nu} \right) \right).
\end{equation}
We start integrating over the $\tau$ variables, and we find
\begin{equation}
Z_I=\sum_{\sigma}\exp\Big(\frac{\beta}{2N}\sum_{ij}^N(\sum_{\nu}^{\alpha N} \xi_i^{\nu}\xi_j^{\nu})\sigma_i\sigma_j \Big)\int \prod_{{\mu}=1}^P d\mu(z_{\mu})
\exp\Big( \sum_{\mu}^{\gamma N} z_{\mu}\Phi_{\mu}+ \epsilon^2 \sum_{\mu \mu'}^{\gamma N} z_{\mu} \Psi_{\mu \mu'} z_{\mu'} \Big),
\end{equation}
where the effective inputs $\Phi$ and $\Psi$ are given by
\begin{eqnarray}
\Phi_{\mu} &=& \frac{\sqrt{\beta}}{\sqrt{N}}\sum_i \xi_i^{\mu}\sigma_i + \epsilon \frac{\beta}{N}\sum_i \sigma_i (\sum_{\nu} \xi_i^{\nu} \xi_{\mu}^{\nu}), \\
\Psi_{\mu \mu'} &=& \frac{\beta}{2N}(\sum_{\nu} \xi_{\mu}^{\nu} \xi_{\mu'}^{\nu}).
\end{eqnarray}
\newline
Next, we use the mean field approximation by which $\sum_{\mu'}\Psi_{\mu \mu'} z_{\mu'} \sim -(\alpha \beta^2/2) z_{\mu}$.
Therefore, we can bound the expression above with the partition function
\begin{equation}
Z_I \sim \sum_{\sigma}\exp\Big( \frac{\beta}{2N}\sum_{i j}^N (\sum_{\nu}^{\alpha N} \xi_i^k \xi_j^k + \sum_{\mu}^{\gamma N} \xi_i^{\mu} \xi_j^{\mu} \frac{1}{\sqrt{1+ \epsilon \alpha \beta^2}}) \sigma_i \sigma_j  \Big).
\end{equation}
If we perform the same procedure, integrating first on $z$ and then on $\tau$, we obtain the specular result
\begin{equation}
Z_I \sim \sum_{\sigma}\exp\Big( \frac{\beta}{2N}\sum_{i j}^N (\sum_\mu^{\gamma N} \xi_i^{\mu} \xi_j^{\mu} + \sum_{\nu}^{\alpha N} \xi_i^{\nu} \xi_j^{\nu}\frac{1}{\sqrt{1+ \epsilon \gamma \beta^2}}) \sigma_i \sigma_j  \Big),
\end{equation}
To obtain the final equation for the partition function, we sum the two Hamiltonians and divide by two, to find
$$
Z_I \sim \sum_{\sigma} \exp \left( \frac{\beta}{4N} \sum_{ij}^N \left(  \sum_{\nu}^{\alpha N}\xi_i^{\nu}\xi_j^{\nu} \left(1 + \frac{1}{\sqrt{1+\epsilon \beta^2 \gamma}} \right)  \right) + \sum_{\mu}^{\gamma N}\xi_i^{\mu}\xi_j^{\mu} \left(1 + \frac{1}{\sqrt{1+\epsilon \beta^2 \alpha}} \right)\right).
$$
Retaining only the first order terms in $\epsilon$, we obtain an equivalent Hamiltonian for a HBM where the hidden layers interact.
This  corresponds to a Hopfield model with an additional noise source, characterized by the Hamiltonian
\begin{equation}
H(\sigma;\xi,\eta)=\frac{\beta}{2N}\sum_{ij}^N\Big( \sum_{\nu}^{\alpha N} \xi_i^{\nu}\xi_j^{\nu} [1 - \epsilon \beta^2 \gamma/4] + \sum_{\mu}^{\gamma N} \xi_i^{\mu}\xi_j^{\mu} [1 - \epsilon \beta^2 \alpha/4] \Big).
\end{equation}
Note that for $\epsilon=0$ we recover the standard Hopfield model.
The effect of the additional noise source on retrieval of patterns corresponding to one layer depends on the load of the other layer: the larger number of neurons in one layer, the larger the perturbation on the retrieval of the other layer.

\section{Conclusions}

 We demonstrate an exact mapping between the Hopfield network and a specific type of Boltzmann Machine, the Hybrid Boltzmann Machine (HBM), in which the hidden layer is analog and the visible layer is digital.
This type of structure is novel, since previous studies have investigated the cases in which both types of layers are either analog or digital.
The thermodynamic equivalence demonstrated in our study paves the way to a novel procedure for simulating large Hopfield networks: In particular, Hopfield networks require updating $N$ neurons and storing $N(N-1)/2$ synapses, while HBM require updating $N+P$ neurons and storing only $NP$ synapses, where $P$ is the number of stored patterns.

In addition, the well known phase transition of the Hopfield model has a counterpart in the HBM.
In Boltzmann Machines, the ratio between the sizes of the hidden and visible layers is arbitrary and needs to be adjusted in order to obtain the optimal generative model of the observed data.
If the number of hidden units is too small, the generative model is over-constrained and is not able to learn, while if it is too big then the model "overlearns" (overfits) the observed data and is not able to generalize \cite{benjo}.
Interestingly, these two extrema correspond in the Hopfield model to, respectively, the low storage phase, in which only a few patterns can be represented,
 and the spin glass phase, in which there is an exponentially increasing number of stable states.
Therefore, the corresponding phase transition in the HBM can be understood as the optimal trade-off between flexibility and generality, thus effectively representing a statistical regularization procedure \cite{hotel3}.

Furthermore we showed that, if hidden layers are disconnected, the corresponding patterns contribute linearly to the capacity of the Hopfield network.
Therefore, conditional independence among layers corresponds to linearity of the energy function.
Instead, if the hidden layers interact, we show that they affect retrieval by acting as an effective noise source.
Although the replica trick has represented a breakthrough for studying the thermodynamics of the Hopfield model, we argue that the "natural" mathematical backbone required for studying the thermodynamics of the Boltzmann machine is the stochastic stability, whose implementation is tractable.

Our work further contributes on connecting scientific communities quite far apart, such as the mathematical physicists studying spin glasses (see i.e. \cite{Bovierbook}) and the computer scientists studying machine learning and artificial intelligence (see i.e. \cite{marc}).

\section*{Acknowledgments}

The strategy outlined in this research article belongs to the study supported by the Italian Ministry for Education and Research (FIRB grant number $RBFR08EKEV$) and by Sapienza Universit\`a  di Roma.
\newline
Adriano Barra is partially funded by GNFM (Gruppo Nazionale per la Fisica Matematica).
\newline
Adriano Barra is grateful to Elena Agliari and Francesco Guerra for useful discussions.

\section*{Appendix 1}

In this appendix, we calculate the interpolating free energy $\tilde{A}(\beta, t)$ for $t=0$. 
This calculation involves only one-body terms and is equal to
\begin{equation}\nonumber
\tilde{A}(\beta,t=0) = \frac{\mathbb{E}}{N}\log \sum_{\sigma} \int \prod_{\mu=1}^P d\mu(z_{\mu}) \int \prod_{\nu=1}^K d\mu(\tau_{\nu})e^{a\sum_i^N \tilde{\eta}_i\sigma_i  + b \sum_{\mu=1}^P \tilde{\eta}_{\mu}z_{\mu} + c \sum_{\nu}^K \tilde{\eta}_{\nu}\tau_{\nu}}e^{\frac{h}{2}\sum_{\mu=1}^P z_{\mu}^2 + \frac{\epsilon}{2} \sum_{\nu=1}^K \tau_{\nu}^2}.
\end{equation}
Due to the additive properties of the logarithm (i.e. $\log(A\cdot B \cdot C) = \log A + \log B + \log C$) and the one-body factorization within each layer, the equation above can be rewritten as a sum of three separate terms, one for each layer:
\begin{eqnarray}
\tilde{A}(\beta,t=0)
&=& \frac{\mathbb{E}}{N} \log \sum_{\sigma}  e^{a\sum_i^N \tilde{\eta}_i\sigma_i } + \\
&=& \frac{\mathbb{E}}{N} \log \int \prod_{\mu=1}^P d\mu(z_{\mu})e^{b \sum_{\mu=1}^P \tilde{\eta}_{\mu}z_{\mu}} e^{\frac{h}{2}\sum_{\mu=1}^P z_{\mu}^2}\\
&=& \frac{\mathbb{E}}{N} \log \int \prod_{\nu=1}^K d\mu(\tau_{\nu}) e^{c \sum_{\nu}^K \tilde{\eta}_{\nu}\tau_{\nu}}
e^{ \frac{\epsilon}{2} \sum_{\nu=1}^K \tau_{\nu}^2}.
\end{eqnarray}
We show in Appendix $2$ shows that the following choice of the parameters substantially simplifies the calculations, $ a=\sqrt{\beta(\alpha\bar{p}+\gamma\bar{r})}, \qquad b=\sqrt{\beta\bar{q}}, \qquad c=\sqrt{\beta\bar{q}},\qquad h=\epsilon=\beta (1-\bar{q})$.
Using these values of the parameters, and performing the integrals and sums in the above expression, we find
\begin{eqnarray}\label{interpo0}
\tilde{A}(\beta,t=0)&=&\log2+\int d\mu(\eta)\log\cosh(\sqrt{\beta(\alpha\bar{p}+\gamma\bar{r})}\eta)+\frac{\alpha+\gamma}{2}\log(1-\beta(1-\bar{q}))^{-1}\nonumber\\
&+&\frac{\beta(\alpha+\gamma)}{2}\frac{\bar{q}}{1-\beta(1-\bar{q})}.
\end{eqnarray}

\section*{Appendix 2}

In this section we focus on the $t$-derivative of $\tilde{A}(\beta, t)$.
Since the interpolating parameter $t$ appears seven times in the exponential, this derivative includes seven different terms.
Their derivation is long but straightforward, here we report the result for each of the seven terms
\begin{align}
&-\frac{\alpha\beta}{2}\langle q_{12}p_{12}\rangle+\frac{\beta}{2N}\mathbb{E}\sum_{\mu}\omega(z_{\mu}^2),\\
&-\frac{\gamma\beta}{2}\langle q_{12}r_{12}\rangle+\frac{\beta}{2N}\mathbb{E}\sum_{\nu}\omega(\tau_{\nu}^2),\\
&-\frac{a^2}{2}(1-\langle q_{12}\rangle),\\
&\frac{\alpha b^2}{2}\langle p_{12}\rangle-\frac{b^2}{2N}\mathbb{E}\sum_{\mu}\omega(z_{\mu}^2),\\
&\frac{\gamma c^2}{2}\langle r_{12}\rangle-\frac{c^2}{2N}\mathbb{E}\sum_{\nu}\omega(\tau_{\nu}^2),\\
&\frac{h}{2N}\sum_{\mu}\mathbb{E}\sum_{\mu}\omega({z_{\mu}^2}),\\
&\frac{\epsilon}{2N}\sum_{\mu}\mathbb{E}\sum_{\nu}\omega({\tau_{\nu}^2}).
\end{align}
Pasting the various terms together we obtain
\begin{align}
\frac{d\tilde{A}}{dt}=&\mathbb{E}\sum_{\mu}\omega({z_{\mu}^2})\Big(\frac{\beta}{2N}-\frac{b^2}{2N}-\frac{h}{2N}\Big)+\mathbb{E}\sum_{\nu}\omega({\tau_{\nu}^2})\Big(\frac{\beta}{2N}-\frac{c^2}{2N}-\frac{\epsilon}{2N}\Big)-\frac{\alpha\beta}{2}\langle q_{12}p_{12}\rangle\nonumber\\
&-\frac{\gamma\beta}{2}\langle q_{12}r_{12}\rangle-\frac{a^2}{2}(1-\langle q_{12}\rangle)+\frac{\alpha b^{2}}{2}\langle p_{12}\rangle+\frac{\gamma c^{2}}{2}\langle r_{12}\rangle.
\end{align}
We are left with the freedom of choosing the most convenient parameters; we see that with the particular choice
$$
a=\sqrt{\beta(\alpha\bar{p}+\gamma\bar{r})}, \qquad b=\sqrt{\beta\bar{q}}, \qquad
c=\sqrt{\beta\bar{q}},\qquad h=\epsilon=\beta (1-\bar{q}),
$$
we can express the whole derivative as the source of the overlap fluctuations
\begin{equation}\label{strilla}
\frac{d\tilde{A}}{dt} = -\frac{\beta}{2}\langle(q_{12}-\bar{q})[\alpha(p_{12}-\bar{p})
+\gamma(r_{12}-\bar{r})]\rangle +  \frac{\beta}{2}(\bar{q}-1)(\alpha\bar{p}+\gamma\bar{r})-\frac{\beta(\alpha+\gamma)}{2}.
\end{equation}
The first term in the right hand side represents the fluctuations of each order parameter around its average (i.e. $\bar{q},\bar{p},\bar{r}$), and we neglect this term within a replica symmetric approach.
The second term includes only averages and does not depend on $t$. 
Its integration in $t$ on the interval $0,1$ coincides with multiplication by one.

\section*{Appendix 3}
In this section, we calculate the free energy in presence of an external field designed to force retrieval of the stored patterns. 
When the stored patterns are Gaussians, and in the thermodynamic limit, retrieval is not a spontaneous emergent feature of the network.
However, it is possible to force retrieval by adding a proper Lagrange multiplier in the interpolating free energy as $t m_1^2 + (1-t)m_1 M_1$, where $m_1= N^{-1}\sum_i^N \xi_i^1 \sigma_i$ is the Mattis magnetization of the first condensed pattern (we chose the first because there is full permutational invariance among patterns) and $M_1$ is its replica symmetric approximation.

In analogy with the calculation performed in Section $4$, we find the following expression
\begin{align}
&\frac{1}{N}\mathbb{E}(\log Z(\beta,\xi,\eta))+\frac{\beta}{2}\int_{0}^{1}dt\langle (q_{12}-\bar{q})[\alpha (p_{12}-\bar{p})+\gamma (r_{12}-\bar{r})]\rangle\nonumber\\
&=\frac{\beta}{2}\int_{0}^{1}dt\langle (m_{1}-M)^{2}\rangle+A^{RS}(\bar{p},\bar{q},\bar{r},M;\alpha,\beta,\gamma),
\end{align}
Fluctuations of $m_1$ around $M_1$ are now present.
The final replica symmetric free energy can be written as
\begin{eqnarray}\label{RSFE}
A^{RS}(\bar{p},\bar{q},\bar{r},M;\alpha,\beta,\gamma) &=&\log 2+\int d\mu(\eta)\log\cosh\Big(\eta\sqrt{\beta(\alpha\bar{p}+\gamma\bar{r})+\beta^{2}M^{2}}\Big)\nonumber\\
&+&\frac{\alpha+\gamma}{2}\log \Big(\frac{1}{1-\beta (1-\bar{q})}\Big)+\frac{(\alpha+\gamma)\beta}{2}\frac{\bar{q}}{1-\beta (1-\bar{q})}\nonumber\\
&-&\frac{\beta}{2}(\alpha\bar{p}+\gamma\bar{r})(1-\bar{q})-\frac{(\alpha+\gamma)\beta}{2}-\frac{\beta}{2}M^2.
\end{eqnarray}
We have to minimize the free energy (\ref{RSFE}) with respect to the replica symmetric order parameters $\bar{q},\bar{p},\bar{r},M$, namely we impose that
\begin{equation}\nonumber
\partial_{\bar{q}}A^{RS}(\beta;\alpha,\gamma)=0,\qquad\partial_{\bar{p}}A^{RS}(\beta;\alpha,\gamma)=0,
\qquad\partial_{\bar{r}}A^{RS}(\beta;\alpha,\gamma)=0,\qquad\partial_M A^{RS}(\beta;\alpha,\gamma)=0.
\end{equation}
This gives the following system of integrodifferential equations to be simultaneously satisfied
\begin{align}
&\partial_{\bar{q}}A^{RS}=\frac{\beta}{2}\Big(\alpha\bar{p}+\gamma\bar{r}-\frac{(\alpha+\gamma)
\bar{q}\beta}{(1-\beta(1-\bar{q}))^{2}}\Big)=0,\\
&\partial_{\bar{p}}A^{RS}=\frac{\alpha\beta}{2}\Big(\bar{q}-\int d\mu(\eta)\tanh^{2}\Big(\eta\sqrt{\beta(\alpha\bar{p}+\gamma\bar{r})+\beta^{2}M^{2}}\Big)=0, \\
&\partial_{\bar{r}}A^{RS}=\frac{\gamma\beta}{2}\Big(\bar{q}-\int d\mu(\eta)\tanh^{2}\Big(\eta\sqrt{\beta(\alpha\bar{p}+\gamma\bar{r})+\beta^{2}M^{2}}\Big)=0,\\
& \partial_M A^{RS}= \int d\mu(\eta) \tanh\Big(\eta\sqrt{\beta(\alpha\bar{p}+\gamma\bar{r})+\beta^{2}M^{2}}\Big),
\end{align}
which can be solved numerically.

\end{document}